\documentclass[12pt]{article}
\usepackage{fullpage}
 \usepackage{url}
 \usepackage{graphicx}
\usepackage{lineno}
\usepackage{epstopdf}
\usepackage{caption}
\newcommand{\Unio}{Unionidae}
\newcommand{\Rebekah}{Rebek\mbox{}ah }
\newcommand{\Cgig}{\emph{C. gigas}}
\newcommand{\CgigFull}{\emph{Crassostrea gigas}}

\newcommand{\Espin}{\emph{E. spinosa}}
\newcommand{\EspinFull}{\emph{Elliptio spinosa}}
\newcommand{\Ecras}{\emph{E. crassidens}}
\newcommand{\EcrasFull}{\emph{Elliptio crassidens}}

\usepackage[ocgcolorlinks,citecolor=blue]{hyperref}
\usepackage{lineno}
 \usepackage{setspace}
       
\begin{document}

\author{\small\Rebekah L. Rogers$^{1*}$, John P. Wares$^{2,3}$, and Jeffrey T Garner$^{4}$}

\title{Genome evolution in an endangered freshwater mussel}
\date{}

\maketitle
\noindent1.  Dept of Bioinformatics and Genomics, University of North Carolina, Charlotte, North Carolina \\
\noindent2.  Dept of Genetics, University of Georgia, Athens, Georgia \\
\noindent3.  Georgia Museum of Natural History, Athens, Georgia \\
\noindent4. Division of Wildlife and Freshwater Fisheries, Alabama Department of Conservation and Natural Resources, Florence, Alabama \\

\noindent*  Corresponding Author: Rebek\mbox{}ah.Rogers@charlotte.edu \\


\noindent Short Title: Genomics in an endangered freshwater mussel \\

 \clearpage
\subsubsection*{Significance Statement}
Nearly-neutral theory predicts that evolution will operate differently in small populations. Adaptive change is less common, and detrimental variation may accumulate.  We test these theories with a new reference genome for a critically endangered freshwater mussel.  Results for SNPs and duplications are fully consistent with nearly neutral theory. However, transposable elements do not follow expected rules suggesting a more complex interplay between mutation, selection, and drift.  This study alters our understanding of nearly neutral evolution, but raises new questions about selfish genetic elements.  Results will influence conservation genomics as it offers a clear case study for how evolutionary studies of gene duplications and transposable elements can help us understand forces that have shaped endangered populations in the recent past. 

\clearpage

\subsubsection*{Abstract}
Nearly neutral theory predicts that evolutionary processes will differ in small populations compared to large populations, a key point of concern for endangered species.  The nearly-neutral threshold, the span of neutral variation, and the adaptive potential from new mutations all differ depending on $N_e$.  To determine how genomes respond in small populations, we have created a reference genome for a US federally endangered IUCN Red List freshwater mussel, \emph{Elliptio spinosa}, and compare it to genetic variation for a common and successful relative, \emph{Elliptio crassidens}.  We find higher rates of background duplication rates in \emph{E. spinosa} consistent with proposed theories of duplicate gene accumulation according to nearly-neutral processes.  Along with these changes we observe fewer cases of adaptive gene family amplification in this endangered species.  However, TE content is not consistent with nearly-neutral theory.  We observe substantially less recent TE proliferation in the endangered species with over 500 Mb of newly copied TEs in \emph{Elliptio crassidens}.  These results suggest a more complex interplay between TEs and duplicate genes than previously proposed for small populations.  They further suggest that TEs and duplications require greater attention in surveys of genomic health for endangered species.  

\clearpage
\subsection*{Introduction}
Nearly-neutral theory predicts that in large populations evolutionary processes work differently compared to evolution in small populations. When population sizes are small, forces of genetic drift may overwhelm the forces of natural selection, resulting in reduced instances of adaptation. The threshold at which mutations behave as though neutral equivalent depends directly on the effective population size.  So long as $4N_e s >1$, selection rules the day to weed out detrimental variation \cite{Kimura1983,Ohta1973} and cause beneficial alleles to spread efficiently in deterministic selective sweeps \cite{haldane1927,Maynard1971,Hermisson2005,Gillespie1991} (where $N_e$ is the effective population size and $s$ is the selection coefficient).  When $4N_e s < 1$, forces of drift can allow mildly detrimental variation to spread under stochastic fluctuations \cite{Kimura1983,Ohta1973}.  Hence, the threshold of variation that will be tolerated as nearly neutral polymorphism is expected to shift when population sizes drop \cite{Ohta1973}.  Under such circumstances, detrimental variation is expected to accumulate in small, isolated populations.

Beyond the nearly neutral threshold, evolutionary potential that can contribute to adaptive changes is expected to be weaker in small populations.  The amount of standing variation in natural populations scales according to $\theta = 4 N_e \mu$, where $\mu$ is the mutation rate \cite{Watterson1975,Tajima1983}. Hence large populations will carry a greater spectrum of genetic variation that would be immediately available if selective pressures shift. If necessary mutations for adaptation are not present among standing variation when new selective pressures arise \cite{Hermisson2005}, then the timeline to generate new variation that establishes a deterministic selective sweep \cite{haldane1927} also depends on $N_e$ according to $T_e = \frac{1}{4N_e s}$ \cite{Maynard1971}, so that large populations will have shorter wait times than small populations.   Collectively, these theoretical models predict that adaptive changes will be rarer in small populations than in large populations. 

Nearly neutral models were initially envisioned with an eye toward single basepair changes in DNA \cite{Kimura1983,Ohta1973}, but modern evolutionary genomics has expanded the predictions to encompass new expectations for mutations that modify larger segments of DNA at once.  Gene duplications are expected on average to be neutral or detrimental resulting in their accumulation in smaller populations \cite{LynchBook, Lynch2000}.  In addition, non-adaptive subfunctionalization is expected to be more common in small population sizes \cite{Lynch2000b}.  Hence, total genomewide duplication rates are predicted to increase as population size decreases \cite{LynchBook, Lynch2000,Lynch2000b}.  Alongside this increase in effective mutation rates, adaptive duplication will be less common and will spread through populations less efficiently as population sizes drop \cite{haldane1927,Maynard1971}.  Similarly, transposable elements that copy at the expense of host genomes are a known source of detrimental variation that is predicted to accumulate in species with smaller population sizes \cite{LynchBook}.  These genomic features have clear predictions from modern nearly neutral theory, and suggestive evidence in cross-phyla comparisons.  In the modern genomic era, it is now possible to explore the empirical outcomes of these nearly-neutral theories using a whole genome analysis of SNPs, duplications and TEs in closely related species with large differences in population size.  

Freshwater mussels (Family Unionidae) have experienced recent population size changes across multiple species, with over 70\% of the approximately 300 species in North America threatened, endangered or extinct \cite{Strayer2004,Haag2014,Garner2001}.  However, life history strategies and habitat requirements among individual species vary greatly so the results of these factors have been highly disparate among freshwater mussels \cite{Haag2014,Haag2013, Garner2001}.  Among these \emph{Elliptio spinosa} (Lea, 1836) is critically endangered \cite{us2010endangered,us2011endangered,Cummings2012} and has not been seen in the wild since 2011 \cite{FiveYr}.  We fortuitously gained the opportunity to pursue whole genome sequencing for \emph{E. spinosa} from DNA collected at Coon Island, Georgia, USA using non-destructive hemolymph sampling in 2007 \cite{Small2009}.  Alongside this specimen, hDNA available from museum collections sampled in the 1980s and later allowed us to compare this individual to diversity present in the population at earlier timepoints.  To determine how evolutionary outcomes might differ in this small endangered population, we compare genomic features to a new reference genome and annotations generated for a close relative in the same genus, \emph{Elliptio crassidens} (Lamarck, 1819), with broad species range and very large census sizes considered ``Currently Stable" \cite{Williams1993,Williams2008}.  These questions about the impacts of drift and inbreeding have been key concerns for conservation, but with little available genetic data to inform management \cite{FiveYr}.

This unique opportunity to build new genome resources for a critically imperiled IUCN Red List \cite{Cummings2012} and US Federally Endangered Species \cite{us2010endangered,us2011endangered,FWS} allows us to examine the impacts of genetic variation and the evolutionary forces that have shaped diversity even when living individuals are difficult to find in the wild.  Using computational bioinformatic techniques and scans of selection, we can identify single base pair differences, rapidly evolving gene families, duplicate genes that are targets of selection, and TE proliferation in species that have multiple orders of magnitude differences in population sizes.  With these new genomic resources we can use these freshwater mussels as a model and case study for empirical outcomes of nearly neutral evolution in TEs and duplicate genes. 
The great imperilment of North American freshwater mussels, as well as those in Europe \cite{Ollard2023,Lopes2023,Strayer2004}. were brought about by habitat destruction, pollution, barriers along rivers, and loss of fish host species. These anthropogenic changes represent critical selective pressures on Unionid mussels \cite{Strayer2004,Haag2014,woodside2004,ortmann1924,savitz1996}.  The end result is that some species have experienced extreme population declines, while others retain massive populations.    As new genomes emerge for this clade, we can use these molluscs to determine how SNPs, duplications, and TEs respond in large and small populations during rapid and drastic shifts in selective pressures under anthropogenic change.  Here our genomic analysis offers the chance to define empirical outcomes of nearly neutral processes while simultaneously offering guidance about endangered clades that may be useful in population management and real world strategies for conservation.

\subsection*{Results}

\subsubsection*{Genome Size and Gene Content}
We assembled a new draft reference genome for \Espin {} that is 2.86 GB with an N50=285,217 and a reference genome for \Ecras {} that is 3.49 GB with an N50=232,187.  Assembly quality judged from BUSCO completeness \cite{simao2015,manni2021} is very similar with 92.3\% complete BUSCOs and 3.3\% fragmented BUSCOs in \Ecras {} and 92.6\% complete BUSCOs and 3.4\% fragmented BUSCOs in \Espin. In \emph{E. spinosa} we find  32,679 putative genes and 33,630 transcripts and in\emph{E. crassidens} 34,673 genes with 39,964 transcripts (Table \ref{Annot}).  Isoform prediction in \emph{E. crassidens} is expected to have better power given RNAseq data available with matched species.  Gene length is not significantly different between the two species (Wilcox Rank Sum Test $W = 571161626$, $P = 0.06683$) with a mean gene length of 17,070 in \Espin {} and 18,648 in \Ecras.  However, intron length is significantly longer in \Ecras {} than in \Espin {} ($W = 1.2638\times10^{10}$, $P=0.0002937$).  Predicted coding sequences (CDS) are longer in \Espin {} than \Ecras {} ($W = 769745873$, $P < 2.2\times 10^{-16}$).  These results suggest that genes are more compact in \Espin {} than \Ecras.  We identify 13,604 unfiltered one-to-one orthologs in a reciprocal best hit BLAST \cite{BLAST} between \emph{E. spinosa} and \emph{E. crassidens}.

  
  

\subsubsection*{Heterozygosity and Evolutionary constraint}
Genetic diversity is predicted to vary with effective population size, according to $\theta=4N_e\mu$ \cite{Watterson1975}.  At equilibrium, heterozygosity is predicted to be equal to $\theta$ such that $\hat{H}=4N_e\mu$, however under declining population sizes $H_t=(1-\frac{1}{2N})^t H_0$ \cite{Wright1931}.  Given these theories, we predict reduced heterozygosity in small and declining populations compared with large populations \cite{Wright1949}. Severe reductions in genetic diversity can damage population health and chances at species survival, but few genetic datasets were previously available to assess impacts in this endangered species, \emph{E. spinosa} \cite{us2011endangered,FiveYr}. In \emph{Elliptio}, \emph{E. crassidens} currently has large census population sizes and a broad geographic range east of the Mississippi from Florida to Quebec and Ontario.  In contrast, \emph{E. spinosa} is an endemic species previously found only in the Altamaha River and Ohoopee River systems \cite{Keferl1981} but with no living members identified since 2011 \cite{FiveYr}.  Field surveys indicate clear population declines since the 1940s \cite{Wisniewski2005,FWS,FiveYr}. 

\begin{figure}
\begin{center}
\includegraphics[scale=0.5]{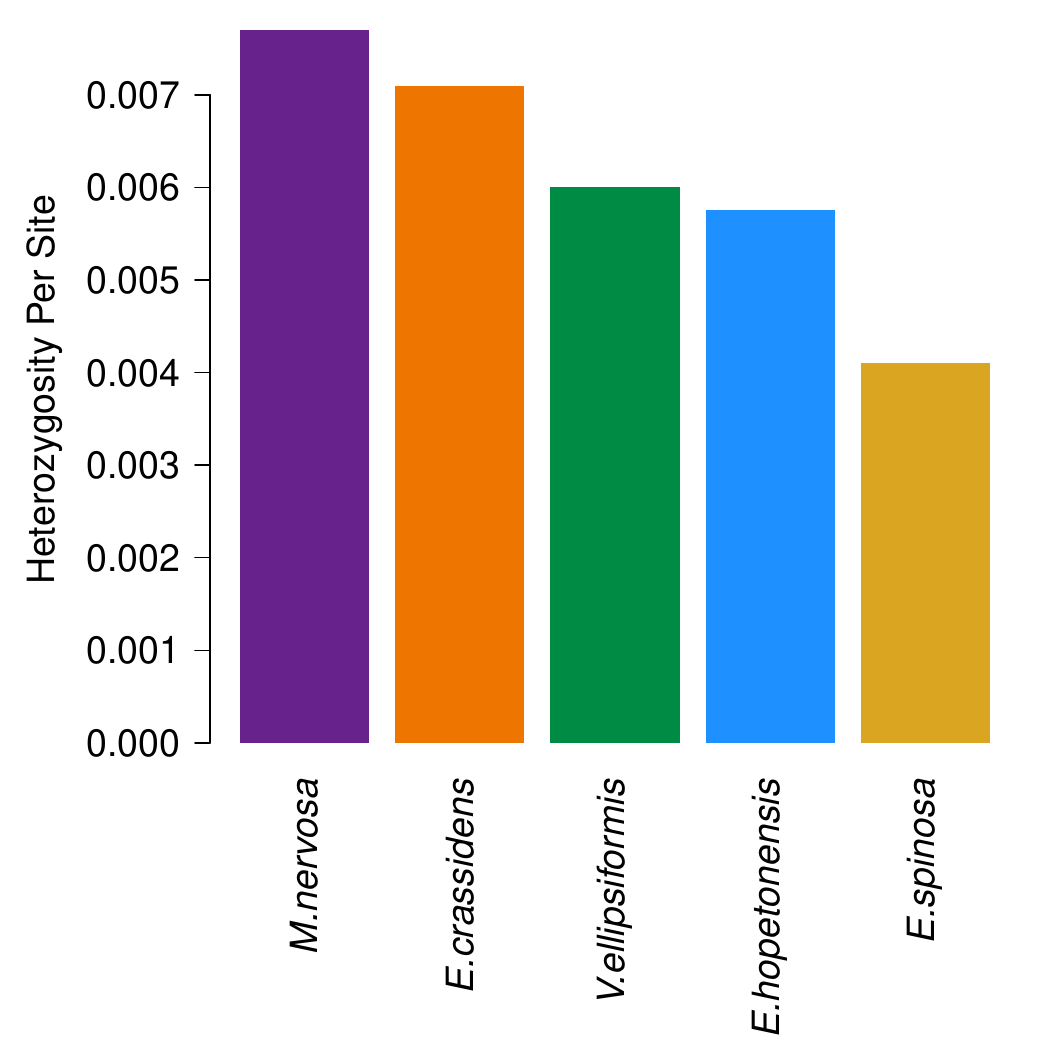}
\caption{Heterozygosity in Freshwater Mussels.  Heterozygosity is high, ranging from 0.0041 to 0.0077 across species.  \emph{Elliptio spinosa}, an IUCN Red List member and US Federally endangered species has the lowest heterozygosity of any Unionid genome sequenced to date.  Error bars are less than the rounding error. \label{BivalveHet} }
\end{center}
\end{figure}

We compare heterozygosity genomewide in 5 species of freshwater mussels where sequence data and genome assemblies are available: \emph{Megalonaias nervosa}  (Rafinesque, 1820) \cite{Rogers2021},  \emph{Venustaconcha ellipsiformis}  (Conrad, 1836) \cite{Renaut2018}, \emph{Elliptio hopetonensis} (Lea, 1838) \cite{Rogers2021}, and two new reference genomes presented here for \emph{Elliptio crassidens} and \emph{Elliptio spinosa}.   \emph{E. spinosa} displays the lowest level of heterozygosity in any of these freshwater mussels with $\theta=0.0041\pm0.000005$.  Correcting for inbreeding by dividing by the portion of the genome that is fully outbred, we estimate $\theta=0.0054$ consistent with its smaller geographic range and available data on prevalence (Figure \ref{BivalveHet}).  Adjusted for inbreeding, genetic diversity is roughly equivalent to \emph{E. hopetonensis}, a non-endangered relative endemic to the Altamaha River.   In hDNA samples NC113372 and NC30078 we estimate heterozygosity of $\theta=0.0139$ and $\theta=0.0125$, respectively. It is difficult to discern how many generations have lapsed since the 1980s in this species.  However if 5 generations have passed estimates could be as low as single digit effective population sizes (n=6-7) and if 10 generations, slightly higher (n=12-13). Some 64 individuals were found across the 1990s \cite{FWS} but no more than 11 at a time in 1993 (Georgia Museum of Natural History record ID 9220). If multiple isolated populations exist along the river \cite{FWS}, global population diversity might be higher than within single demes.  If heterozygosity is inflated by DNA damage (in spite of ethanol preservation and freezing) then population sizes could be lower than genetic estimates.  The \emph{E. spinosa} reference specimen, in contrast, was taken from a fresh hemolymph specimen and maintained at -80C until sequencing, and is therefore far more reliable as an estimate of $N_e$.  \emph{E. crassidens} has the second highest with $\theta=0.0071$ (Figure \ref{BivalveHet}, Table \ref{HetTable}). Still, \emph{E. spinosa} houses fairly robust genetic diversity. Hence, it does not appear that populations are genetically impoverished under recent declines in spite of very small census sizes. The short timescale for population losses \cite{FWS,Wisniewski2005,FiveYr} likely contribute to this observed genetic robustness as few generations have passed since demographic declines.


Under nearly-neutral expectations, we anticipate reduced heterozygosity and permissive heterozygosity for amino acid changing mutations in the smallest populations.  To determine how selective constraint is operating in \emph{Elliptio}, we used the two new contiguous assemblies and annotations for \emph{E. crassidens} and \emph{E. spinosa}, using only those 14,093 genes that have direct orthologs in both species.  We observe reduced selective constraint in the \emph{E. spinosa} reference genome of $H_N/H_S=0.40$, and museum specimens with $H_N/H_S=0.44$ and $H_N/H_S=0.53$ (Table \ref{HetTable}).  In comparison, \emph{E. crassidens} has $H_N/H_S=0.322$ (Table \ref{HetTable}), a difference of 25\% more amino acid substitutions tolerated under small population sizes. Along with this altered selective constraint, we identify a 28\% excess in genes with heterozygous premature stop codons in \emph{E. spinosa} compared with \emph{E. crassidens}, after adjusting for differences in mean heterozygosity (1397 vs 1904, corrected to 2573 vs 1904).  These observations are consistent with nearly-neutral expectations for how single nucleotide diversity would shift in small populations compared with large populations. 


\subsubsection*{Duplicate Gene Evolution}
Nearly-neutral theories of duplicate gene formation have suggested that whole genome background rates of duplication should increase in small populations \cite{Lynch2000,LynchBook}. Because most duplications are neutral or detrimental, mutation rates may escalate in small populations due to reduced selective constraints.  Additionally, non-adaptive subfunctionalization may be more likely in small populations, resulting in greater duplicate gene retention.  We can use birth-death analyses \cite{Lynch2000} to estimate genomewide duplication rates for each species, and to identify rapidly proliferating gene families beyond the bounds of expectations given these genomewide mutations rates. In \emph{E. crassidens} we estimate a whole genome background duplication rate of $2.5\times10^{-8}$ per gene per generation.  In \emph{E. spinosa} (Table \ref{DupRateEspin}), whole genome background duplication rates are 46\% higher than \emph{E. crassidens}, with $3.7\times10^{-8}$ per gene per generation (Table \ref{DupRateEcras}).  

\begin{figure}
A) \includegraphics[scale=0.4]{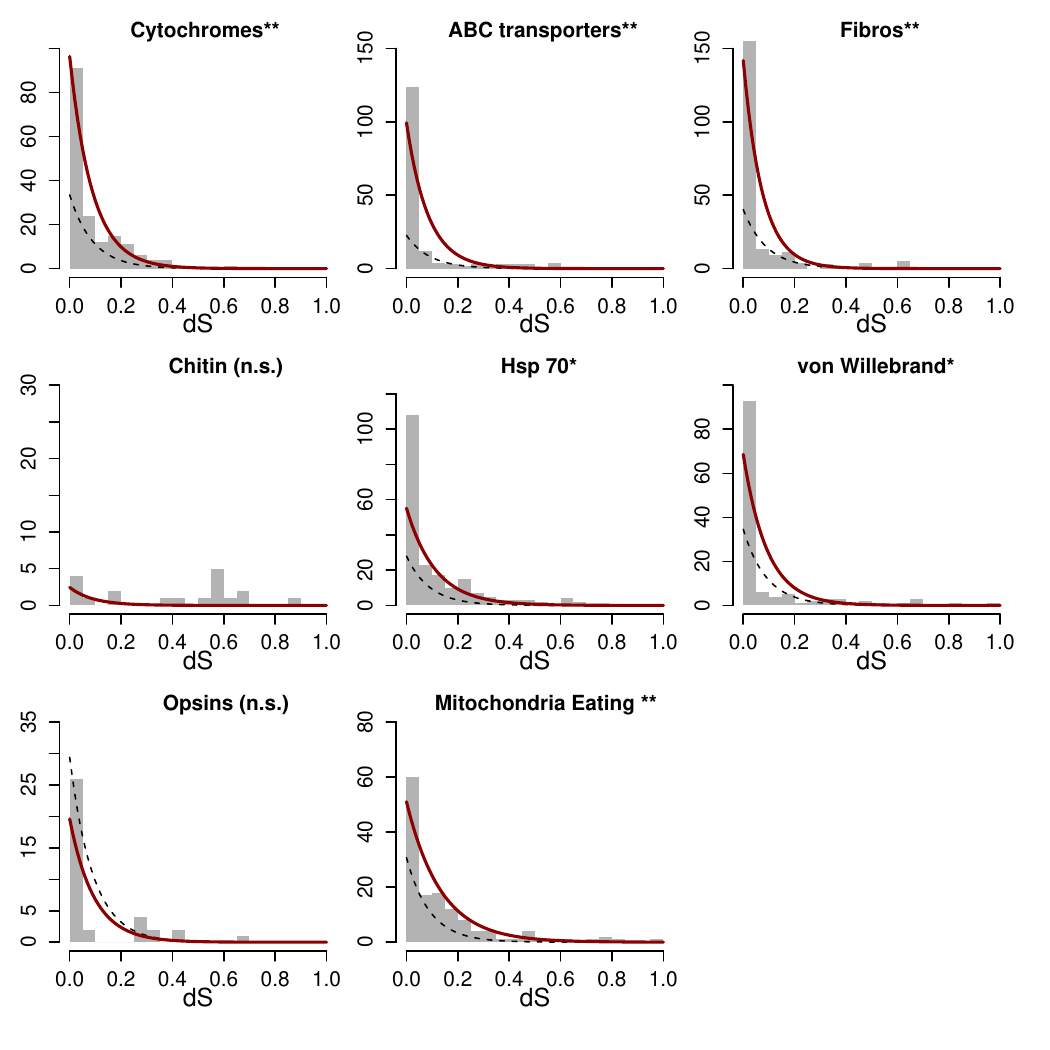}
B) \includegraphics[scale=0.4]{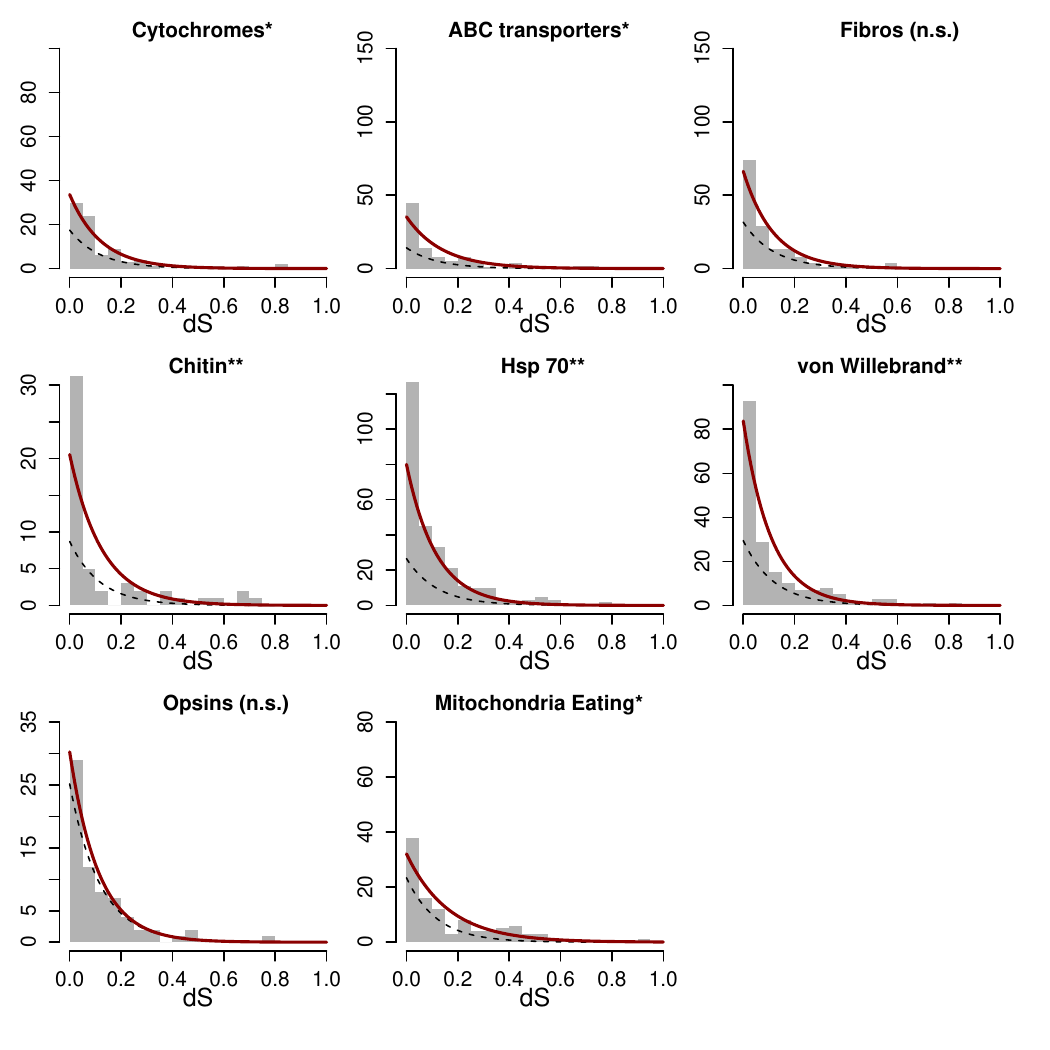}
\caption{Gene Family Amplification in A) \emph{E. spinosa} and B) \emph{E. crassidens}.  We observe greater instances of adaptive gene family amplification in \emph{E. crassidens} for key gene families related tor organism success (red) compared to genomewide background (dashed line).  Cytochromes, ABC transporters, and Heat shock genes are amplified in both species, as are mitochondria management proteins.  We note fibrinogens are more highly amplified in \emph{E. spinosa} while von Willebrand proteins are amplified in \emph{E. crassidens}, suggesting parallel evolution of clotting factors through different pathways in these two species. Opsins also show key differences with few copies and low rates of duplication in \emph{E. spinosa}, consistent with reduced reliance on light queues in a burrowing mussel. \label{GenFamAmp}}
\end{figure}

We identified rapidly evolving gene families in functional classes known to be associated with adaptive duplication in \emph{M. nervosa} \cite{Rogers2021,Rogers2023}.  Mitochondrial management proteins, von Willebrand proteins, heat shock proteins, cytochromes, ABC transporters, all show signatures of rapid amplification, as do Fibrinogens, an alternative pathway to anticoagulation that appears to be especially important in \emph{E. spinosa} but was not identified in prior work on \emph{M. nervosa} (Figure \ref{GenFamAmp}, Table \ref{DupRateEspin}) \cite{Rogers2021}.  The observed parallelism in duplication for anticoagulation pathways rather than strict functional convergence using only von Willebrand factors could reflect use of different fish hosts across different mollusc species. We identify rapid amplification in opsins and chitin metabolism genes important for shell formation in \emph{E. crassidens} but not in \emph{E. spinosa} (Figure \ref{GenFamAmp}, Table \ref{DupRateEspin}-\ref{DupRateEcras}).  \emph{E. spinosa} burrows into sediment up to 10 cm \cite{FWS, Johnson1970}, and emerges during reproduction, potentially reducing the need for light sensation in this species.  It is unclear the exact causes of chitin amplification in \emph{E. crassidens} and \emph{M. nervosa} \cite{Rogers2021,Rogers2023} but it is possible that protective spikes in \emph{E. spinosa} are an alternative defense against predation, rather than protection through thickening shells. 

\begin{figure}

\includegraphics[scale=0.7]{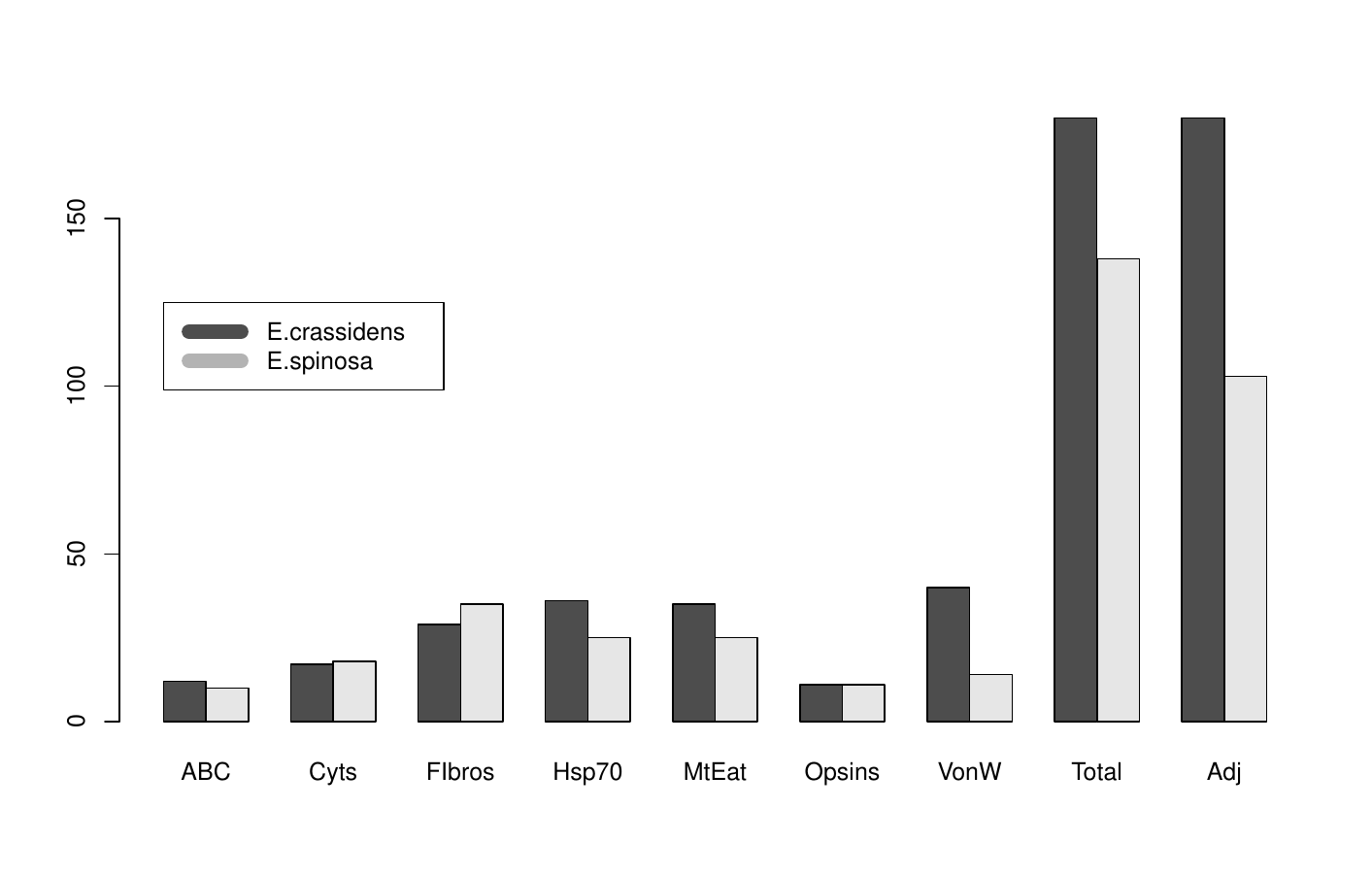}
\caption{Number of Adaptive Duplications in \emph{Elliptio} species.  We observe greater instances of adaptive gene duplication in \emph{E. crassidens} compared to \emph{E. spinosa}, consistent with expectations under nearly-neutral theory.  After adjusting for genomewide duplication rates, differences are even more profound.  Fibrinogens are the exception where \emph{E. spinosa} has greater adaptive amplification of these clotting factors, but far fewer von Willebrand proteins, suggesting different fish-host interactions in these two species.    \label{DupSelnFig}}
\end{figure}

Alongside the increase in duplication mutation rates, these theories also predict that adaptive gene duplication should be less common in small populations where standing variation is reduced, and  deterministic selective sweeps are more difficult to achieve in the face of enhanced genetic drift \cite{haldane1927,Maynard1971}.  Under these circumstances, we expect to observe fewer signals of selection on duplicate genes in small populations. To assess these theories in genomic data for \emph{Elliptio}, we identified gene duplication events with signatures of selection using PAML's codeml package \cite{PAML}.  We used sequences of large gene families within a single species to generate in-frame alignments and estimated dN/dS.  Genes with $dN/dS>1.0$ on the gene tree branch after duplication are considered evidence of positive selection.  We observe a total of 138  duplicate gene pairs with $dN/dS > 1.0$ in \emph{E. spinosa} compared to a total of  180 in \emph{E. crassidens}. If we adjust for the higher background genomewide duplication rates in \emph{E. spinosa}, this would be equivalent to a difference of 103 vs 180 adaptive duplication events, a ratio of 57\% as many adaptive gene duplications in the endangered species (Figure \ref{DupSelnFig}, Table \ref{DupSelnTable}).  Such results are consistent with reduced efficiency of selection as well as the lower adaptive potential expected in very small populations of \emph{E. spinosa}.  We do observe adaptive gene duplication across similar functional categories in both species, with selection on detox genes, anticoagulation genes, heat shock genes, mitochondrial management proteins, and a small number of opsins (Figure \ref{DupSelnFig}, Table \ref{DupSelnTable}). However, only 6 genes that are direct 1:1 orthologs across these two species are targets of selection in both, 2 ABC transporters, 2 von Willebrand factors, 1 Fibrinogen, and 1 mitochondria eating protein.   These results point to parallel selective pressures acting on these two species, with convergence across functional category level but very little gene-level convergence.  While the presence of recent species-specific adaptive gene duplication is encouraging, signatures of adaptation that allow these species to survive and reproduce are reduced in the endangered species, less than 2/3 of expectations based on \emph{E. crassidens}.

In these genomic data, every prediction for duplicate gene content under small population sizes \cite{Lynch2000, Lynch2000b, LynchBook} with reduced adaptive potential \cite{Hermisson2005,Maynard1971} appears to be supported in \emph{E. spinosa}.   

\subsubsection*{Transposable Elements}
\begin{center}
\begin{figure}
\begin{center}
\includegraphics[scale=0.5]{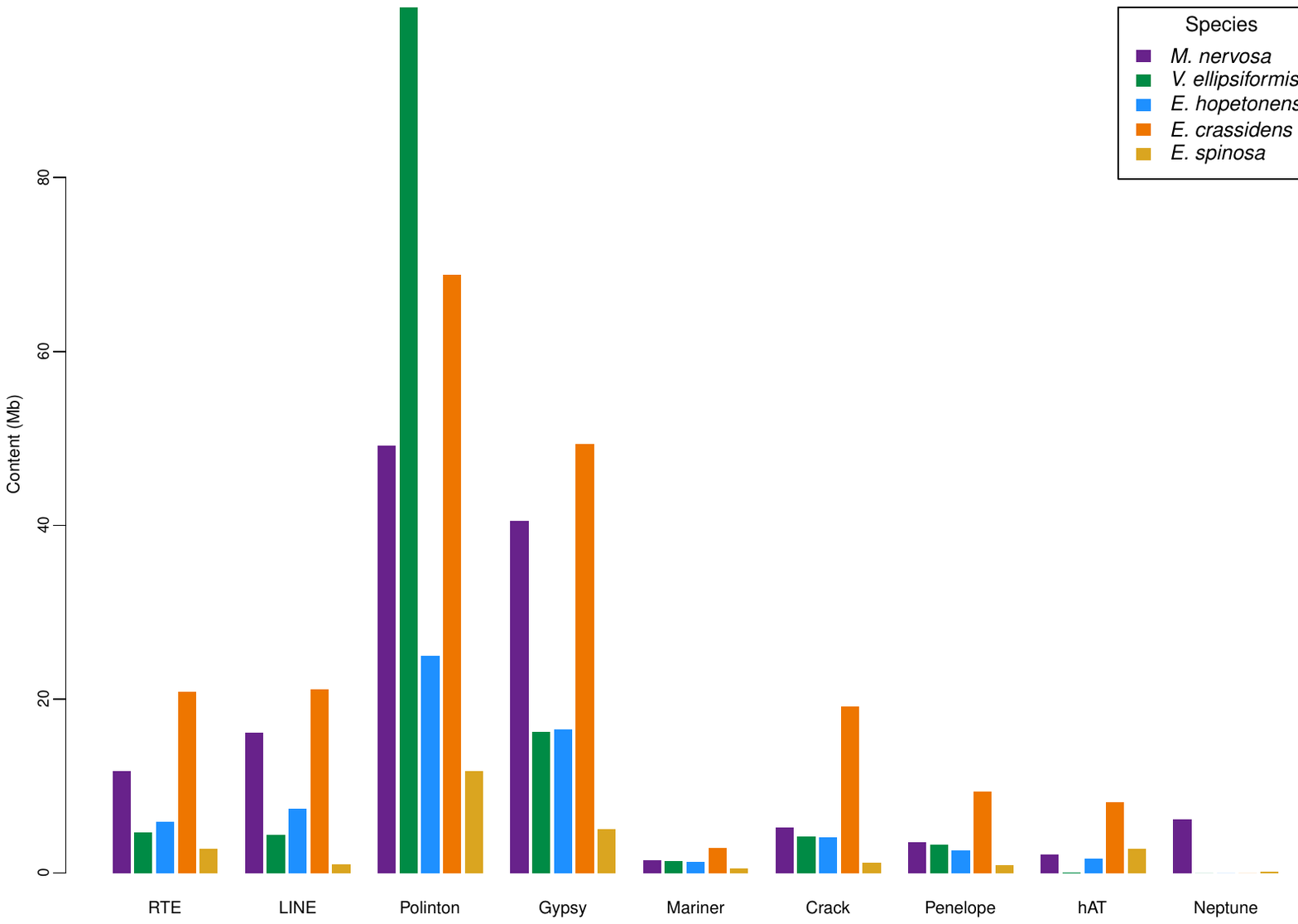}
\caption{Transposable element content in Unionidae genomes.  Total TE content for recently proliferating TEs was estimated using RepDeNovo and mapping reads back to repeat contigs and using coverage to estimate copy number. TEs were identified by family with RepBase database.  Polinton elements are unusually active in \emph{V. ellipsiformis}. \emph{Neptune} elements are present in \emph{M. nervosa} but absent in other species, suggesting putative recent invasion.  \emph{E. crassidens} has higher TE content than \emph{E. spinosa} for every single class of transposable elements, with a total of 610 Mb of excess DNA from recent TE amplification. \label{TEFigure}}
\end{center}
\end{figure}
\end{center}

Transposable elements are expected to induce new mutations that are primarily detrimental \cite{Lynch2003,LynchBook}.  They can fully disrupt gene sequences, alter expression profiles for genes, contribute to new exon formation, and truncate proteins short of full open reading frames \cite{Feschotte2008, Schaack2010, Dubin2018,Bennetzen2000}.  While some individual TE insertions may fortuitously create adaptive variation \cite{Lynch2011,laudencia2012,Schmidt2010}, the presence of active, amplifying TEs is thought to always be maladaptive \cite{Lynch2003,LynchBook,Orgel1980}. Against this backdrop, TE proliferation is predicted to be more common in small populations than large populations \cite{Lynch2003,LynchBook}.  To assess the spectrum of recently active TEs that are most relevant to recent population declines, we identified recently proliferated repetitive elements with less than 2\% sequence divergence across copies using RepDeNovo. We identified TE consensus sequences and copy number in the genomes of 5 freshwater mussels, including \emph{E. spinosa} and \emph{E. crassidens} (Figure \ref{TEFigure},Table \ref{TETable}).  In contradiction with nearly-neutral expectations, we observe far higher TE content among recently active TEs in \emph{E. crassidens} and the lowest TE content in the endangered endemic species, \emph{E. spinosa}. We identify 711 Mb of repetitive elements that have recently proliferated in \emph{E. crassidens}, the most of any freshwater mussel surveyed to date (Figure \ref{TEFigure},Table \ref{TETable}).  In comparison, \emph{E. spinosa} houses only 101 Mb of recently proliferated repeats (Table \ref{TETable}). All transposable element types (LTR retrotransposons, non-LTR retrotransposons, and DNA transposons) appear to be proliferating recently in \emph{E. crassidens} compared with other species (Figure \ref{TEFigure},Table \ref{TETable}).   The recent transposable element expansion in this genome  is consistent with the estimated genome size difference estimating 628 Mb large genome in \emph{E. crassidens}. 

 Across all freshwater mussel species, \emph{Polinton} (aka \emph{Mavrick}) DNA transposable elements appear to be the most common type identified, followed by \emph{Gypsy} LTR retrotransposons and \emph{RTE} non-LTR retrotransposons (Figure \ref{TEFigure},Table \ref{TETable}). We note that recently proliferated \emph{Neptune} elements appear to be present in \emph{M. nervosa} with 6.21 Mb \cite{Rogers2021}, but not in other freshwater mussel species (Figure \ref{TEFigure},Table \ref{TETable}).  We identify large numbers of unclassified repeats, with over 500 Mb in \emph{E. crassidens} and 75 Mb in \emph{E. spinosa} implying the presence of Unionid specific repeats that are not annotated well in current repeat databases.   We observe no significant correlation between current heterozygosity and total TE content ($R^2=0.5162$, $F=3.2$, $df=3$,  $P=0.1716$).   While power may be low across only 5 genomes, we note that the trendline is in direct opposition with nearly-neutral expectations of higher repeat content in small population sizes.
 
Collectively, these results suggest that transposable elements proliferation is likely to be more complex than previous predictions based on on population size. 


\subsubsection*{Inbreeding}
In small populations, inbreeding becomes increasingly likely as short coalescent times lead to shared ancestry and lack of appropriate mates leads to greater likelihood of consanguineous matings.  To identify genomic regions that may be affected by inbreeding, we located regions that were identical by descent with near-zero heterozygosity. Mean diversity across the genome in \emph{E. spinosa}  is $\theta=0.0041$, or 41 heterozygous sites per 10kb (Figure \ref{BivalveHet}, Table \ref{HetTable}).  In \emph{E. crassidens} $\theta=0.0071$ or 71 heterozygous sites per 10kb (Figure \ref{BivalveHet}, Table \ref{HetTable}). We identified 10kb windows with estimates of $\theta<0.0010$, correcting for the number of sites with sufficient coverage to call heterozygotes, as putative IBD tracts. We observe 762 MB out of the 2.86 GB genome lie in putative IBD tracts,  26.7\% of the genome. This level of inbreeding in \emph{E.spinosa} is consistent with half-sibling mating or equivalent parental relatedness in the reference genome specimen.  It would suggest that equivalent whole genome heterozygosity in a non-inbred genome would be roughly $\theta=0.0056$ (i.e. 0.0041/0.733). 

To determine whether continuous inbreeding over time had affected the species decline, we used hDNA for 3 individuals dating from 1984 (NC113372), 1998 (NC25709), and 2004 (NC30078) graciously provided by the North Carolina Museum of Natural Sciences.  Putative IBD tracts affected at most 6\% of the genome. Some 8203/185339 windows (4.4\%) in NC113372 with at least 1000 bp of coverage and 13635/224060 windows (6.0\%) in NC30078 and have estimates of $\theta<0.001$, adjusting for the number of sites that can be assessed with sufficient coverage to call heterozygotes.  The absence of pervasive inbreeding signals in these historic specimens, including one as recent as 2004, suggests IBD tracts are most likely the product of recent consanguineous mating rather than extended multi-generation breeding of less closely related individuals.

\subsubsection*{Estimates of Adaptive Potential}
The ability to adapt to environmental shifts remains a critical concern in endangered species.  Under small population sizes, genetic diversity is expected to decline and beneficial mutations are anticipated to become rare.  To evaluate the limits of adaptive potential in \emph{E. spinosa}, we evaluate outcomes under population genetic models for adaptation. We estimate the probability of adaptation from standing variation ($P_{sgv}$) \cite{Hermisson2005} and the time to establishment of a deterministic sweep $T_e$ \cite{Maynard1971} for SNPs in 5 species of freshwater mussels and for duplications in 4 species of mussels, then evaluate outcomes under demographic decline to small population sizes (Table \ref{PotentialTable}). Based on long standing patterns of genetic diversity in the genome, historic patterns of adaptive potential in \emph{E. spinosa} modestly less than species with large population sizes. \emph{E. spinosa} with historic $N_e=280,000$ shows $P_{sgv}$ 71\% and 78\% than that of \emph{E. crassidens} and \emph{M. nervosa} (Table \ref{PotentialTable}), consistent with its robust patterns of heterozygosity (Figure \ref{BivalveHet}, Table \ref{HetTable}).  $T_e$ is 27\%-37\% longer than \emph{E. crassidens} and \emph{M. nervosa} but comparable to \emph{E. hopetonensis} and \emph{V. ellipsiformis}.  \emph{E. spinosa} has the highest estimated rate of gene duplication, yielding a $P_{sgv}=0.3604$, the highest historical potential in the clade.  For gene duplications, $T_e$ is even shorter than for SNPs, with as few as 245 generations for new mutations to arise and begin to sweep to fixation.

Under demographic declines, $P_{sgv}$ is expected to decline \cite{Hermisson2005} and $T_e$ is expected to become far longer \cite{Maynard1971}.  However, in these small population sizes, theory suggests a far greater impact on the timeline and ability to adapt through new mutation.  Under a theoretical population size crash to $N=1000$, $T_e$ for \emph{E. spinosa} indicates 280x longer to adapt from new mutation for SNPs and duplications.  However, $P_{sgv}$ shows far more modest effects of 49\% as likely to adapt from standing variation for SNPs 54\% as likely to adapt from standing variation for duplications.  We note that $P_{sgv}$ remains above 0.19 for gene duplications.  These models indicate that preservation of genetic diversity among standing variation is likely to play a critical role in organisms' ability to adapt to shifting environments, and that the timeline to loss of diversity is long. However, even short term census size declines can critically impede a population's ability to adapt through new mutation, with profound effects that render new mutations virtually impossible on historic timelines.
\subsection*{Discussion}

\subsubsection*{Nearly Neutral Evolution}
In small populations, genomes are predicted to have lower levels of genetic diversity \cite{Kimura1983}, reduced selective constraint \cite{Ohta1973}, lower rates of adaptation \cite{Hermisson2005},  higher genomewide duplication rates \cite{Lynch2000}, and higher transposable element activity \cite{LynchBook}.  In small populations, selection becomes less efficient, allowing mildly detrimental variation to accumulate, and adaptive changes are less likely to fix.  Alongside these effects of drift, smaller numbers result in a general loss of genetic diversity. To empirically test these theories, we collected new genetic data for a critically endangered species, \emph{E. spinosa}, and \emph{E. crassidens}, a close relative in the same genus which has a wide species range and large population sizes.    

Every expectation of nearly neutral theory is met for duplicate genes and for SNPs.  We observe a higher rate of genomewide background mutation for \emph{E. spinosa} and lower rates of adaptive gene duplication.  We also find reduced genetic diversity for SNPs, even after correcting for inbreeding.  Selective constraint is reduced in \emph{E. spinosa} with a marked 28\% increase in amino acid substitutions that are tolerated based on $H_N/H_S$.  We further identify an effective increase in heterozygous stop codons relative to $H_S$ suggesting higher proportions of SNP variation that is anticipated to break gene functions.  This genome bears every hallmark of nearly-neutral evolution as assessed by multiple independent metrics of genetic variation.

\subsubsection*{Transposable Element Amplification}
Transposable element content emerges as the sole major exception to nearly neutral expectations. Curiously, transposable element content in \emph{E. crassidens} suggests an expansion of recently active TEs, creating an additional 711 Mb in TE content compared to only 101 Mb in \emph{E. spinosa}.  Genome size estimates from assembly reflect this inflated genome size, with a total of 3.49 GB in comparison to 2.86 GB in \emph{E. spinosa}. We observe activity in DNA transposons and retrotransposons, with multiple groups amplyifying across \emph{E. crassidens}. The increase in intron size may be driven by new TE insertions landing within gene regions outside the CDS.  These results run counter to expectations that TE expansion would be more common under small population sizes.  

The forces governing TE content are debated within the field.  While the average TE insertion is most often detrimental, in rare cases TEs can induce variation that is fortuitously adaptive, especially when populations encounter environmental stress \cite{Belyayev2014,laudencia2012,Dubin2018}.  If there were beneficial TE insertions that have not been silenced, then there could be selective pressures that indirectly result in larger TE content. Some cases of such TE hitchhiking and mutualism are known \cite{Cosby2019,Feschotte2008,Dubin2018}, including in humans \cite{Lynch2011,Singh2023}.  Alternatively it is possible that the current scenario simply represents non-equilibrium dynamics where TEs have evaded cellular repressors, and the organism has not yet evolved new defenses \cite{McLaughlin2017,Cosby2019}.  At present, the lack of chromosome scale genomes is a limitation on identifying where these TEs lie within the genome and how they may interact with specific genes.  However, as genomic resources improve in this clade, opportunities to examine the role of transposons in this natural system will offer new opportunities for how they influence species with large and small population sizes.

\subsubsection*{Genomics for Threatened Organisms}
Over 70\% of freshwater mussels in North America are threatened, endangered, or extinct, and mussels in Europe face parallel population declines \cite{Ollard2023,Lopes2023,Strayer2004}.  Understanding how genomes change in small, endangered populations is a critical component of how their biology has shifted in response to recent anthropogenic changes.  We use computational methods to implement whole genome scans for signatures of selection and rapid evolution.  These whole genome approaches are agnostic with respect to function, and do not rely on human biases to identify candidate genes.  Rather, gene families with significant departures from neutrality reveal those functions that are most important for survival and reproduction.  This type of reverse ecological genetics can reveal aspects of organism biology that human biases might otherwise miss \cite{marmeisse2013}. We identify rapidly evolving gene families that are consistent with the biology of the organism, including some gene families that we would not have identified \emph{a priori}. Detox genes appear to be amplifying rapidly, with ABC transporters and their interaction partners the cytochrome P450 genes showing signatures of selection and rapid amplification in both \emph{E. spinosa} and \emph{E. crassidens}, just as was previously noted for a highly successful mussel \emph{M. nervosa} \cite{Rogers2021,Rogers2023}.  These outcomes are fully consistent with known ecological challenges from pesticides and herbicides as well as other pollutants, and consistent with mechanisms of resistance in other organisms \cite{Schmidt2010,Van2020,Nelson2004}. 

Parallelism means there are multiple solutions to similar but not necessarily identical challenges.  Such results imply that there exist multiple genetic pathways to adaptation, an encouraging prospect for survivorship in freshwater mussels. Fish host interaction genes are also a major genetic response to ongoing selective pressures, consistent with high attrition rates at the parasitic larval stage \cite{Johnson2012,milam2005} and observed hemoraging in glochidia infection \cite{howerth2006}.  However, the Fibrinogens dominate in \emph{E. spinosa}, while von Willebrand factors are most common targets of selection in \emph{E. crassidens}. This functional class was not identified among rapidly evolving gene families in \emph{M. nervosa} \cite{Rogers2021}. It is unknown whether \emph{E. spinosa} uses any similar fish hosts to \emph{E. crassidens} \cite{FWS}, however glochidia infection experiments could not achieve successful maturation on common fish hosts \cite{Johnson2012}.  Given the independent genetic response, wider searches for alternative hosts may be warranted.   However, the use of Fibrinogens as well as von Willebrand proteins and the large number of adaptive duplications suggest that there are multiple genetic mechanisms to respond to challenges of the parasitic lifestyle.  This expanded mutational target increases the prospects for success in responding to selective pressures compared to systems that might require single genes or single target site mutations. Under such a genetic structure we anticipate greater chances of survival, partially mitigating reduced adaptive potential and lower genetic diversity.

\subsubsection*{Potential for Rescue}
Population management and endangered species guidelines for \emph{E. spinosa} raise concerns of reduced genetic diversity and inbreeding based on census sizes alone, but federal agencies have highlighted the lack of genetic data as a limitation in population management \cite{FiveYr,us2011endangered}. Encouragingly, genetic diversity does not show the hallmarks of extreme loss at present, indicating that ancestral diversity remained in the existing population even with small population numbers.   Genetic diversity genomewide outside of inbred regions in \emph{E. spinosa} remains relatively strong for an imperiled species with $\theta=0.0041$ mutations per site.   While bottlenecks and subpopulation isolation  may continue to be cause for concern, this species offers promise for preservation if living individuals can be found.  Severe population declines have occurred in very recent decades since the 1940s \cite{Wisniewski2005,FWS,FiveYr}, and populations likely have had few generations since demographic crashes. Still, even in these small populations, ancestral diversity is likely to be captured at present, though we would expect population variation to decay over time \cite{Wright1931}. At this critical stage, timely action may be necessary for species success, if representative individuals can be located in the wild. Preserving this existing genetic diversity among the population may be essential if opportunities arise to propagate populations in the future. 

Population surveys have raised concerns that any putative remaining populations may be isolated from one another and that also long term effective inbreeding in small populations may be in play \cite{FWS}. In genome sequences for \emph{E. spinosa}, we identify inbreeding tracts that encompass more than 25\% of the genome, a signature that would be expected from consanguineous mating of half siblings.  Museum specimens offering hDNA from the 1980s exclude the possibility of long standing population inbreeding under more distant relatedness but short coalescent times. Here, the use of hDNA from museum specimens helps us distinguish between recent and prolonged inbreeding, with potential implications on population management.  A very small local population comprised of many half siblings or similar close relatives could produces such inbreeding on very short timescales, especially given large clutch sizes in freshwater mussels.  

Reciprocal transplantation \cite{fitzpatrick2020} might protect local populations from sibling-sibling inbreeding that increases the likelihood of recessive detrimental variation that is often the basis of genetic disease \cite{Morton1956,Crow2010,Wright1949,Wright1922}. In these genomes, we may expect forces of purging \cite{garcia2012,kleinman2022} to be very strong, as over 25\% of genes would be affected.  In a species with limited numbers, loss of individuals through genetic inviability and removal of linked genetic variation would only cause additional challenges for a species that has been in serious decline over the past 70 years. If inbreeding can be prevented, the anticipated loss of fitness and death of individuals from potential harmful recessives \cite{Morton1956,Crow2010,Wright1949,Wright1922} might be avoided entirely.
 
\emph{E. spinosa} has not been seen in the wild since 2011 \cite{FiveYr}.  \emph{E. spinosa} prefers muddy substrates and burrows into the sediment for much of its lifetime.  Given this lifestyle niche, it is far more difficult to find in active field work than other species of Unionidae. Genome assemblies and annotations released here offer the opportunity for new eDNA surveys in the future, where sediment eDNA sequencing or PCR screens might identify habitats where \emph{E. spinosa} individuals may exist where physical surveys have not yet identified living specimens.  If these new genetic tools might uncover indirect evidence of \emph{E. spinosa} in natural habitats, field surveys might more efficiently locate areas with stronger chances of surviving individuals.   With such minimal population sizes, genetic tools may be one useful avenue for new information during population management for Unionidae as well as other threatened clades across the tree of life.  

\subsection*{Methods}
\subsubsection*{Specimen Collection}
Jeff Garner collected one reference genome specimen for \EcrasFull {} from Wheeler Reservoir, Tennessee River.  The sample was shipped to UNC Charlotte for processing.  The mussel was dissected and tissues were flash frozen in liquid nitrogen.  To obtain high quality long molecule genomic DNA, we used phenol-chloroform extraction with isopropanol precipitation.  Extracted DNA was used to create Illumina genome sequencing libraries, Nanopore sequencing libraries, and RNA sequencing.  Frozen tissues for gills, mantle, and adductors were used for Illumina RNAseq libraries. J. Wares and S. Small obtained non-destructive hemolymph samples for a live specimen of \EspinFull {} obtained in 2008 under supervision of US Fish and Wildlife staff.  We used a phenol chloroform extraction to obtain DNA. DNA was used for Illumina genomic sequencing and Nanopore long read sequencing.

\subsubsection*{Genome Assembly}
We assembled and annotated \emph{Elliptio} genomes similarly to previous work on \emph{Megalonaias nervosa} \cite{Rogers2021}. To create reference genomes we used Nanopore long reads, PacBio long reads, and Illumina short read sequencing. For \Ecras, we collected three sequencing runs of Illumina paired-end reads with a 325 bp insert size, and 4 flow cells of Nanopore Promethion data (chemistry 9.5 and 10.4 flow cells).  For \Espin, we collected 2 Illumina sequencing runs and 4 Promethion flow cells.  We used the Nanopore Guppy GPU basecaller with highest accuracy.  Nanopore sequencing was performed without size selection as BluePippen purified samples for other mussels had disappointingly truncated read lengths (as short as 2kb) compared to unfiltered extraction profiles in both Nanopore and PacBio sequencing.  Nanopore data were filtered prior to assembly to remove reads shorter than 2 kb.  We used the Masurca hybrid long read and short read genome assembler \cite{masurca} to generate reference genomes.  We used BUSCO  v5.1.3 and v3.0.2 \cite{simao2015,manni2021} to assess genome completeness and gene representation.  BUSCO v5.1.3 uses metaeuk to identify busco essential genes \cite{manni2021}. This annotation pipeline performs well for conserved sequences, but poorly for lineage specific genes not present in the training data set.  


\subsubsection*{Gene Annotation}
We collected \Ecras {} paired-end RNAseq data for gills, mantle, and adductor muscle. We used Oyster River Protocol \cite{ORP} to \emph{de novo} assemble transcriptome data.  Assembled transcripts were mapped onto each reference genome using BLAT, converted to gff "hints" for use with the Augustus \emph{de novo} genome annotation software.  Homology based annotation programs work well for clades with close relatives that have been sequenced and annotated. However, these types of approaches perform poorly for this clade as they are so distant from other sequenced taxa and so few Unionid species have been annotated \cite{Renaut2018, Rogers2021, Smith2021}. We used the species-specific Augustus training files produced in BUSCO 3.0.2  \cite{simao2015} to guide Augustus annotations, allowing for isoform prediction. We used Interproscan v5.53-87.0 \cite{interpro} to identify conserved functional domains in the resulting protein sequences. 

Putative annotations from Augustus were filtered to remove transposable elements (TEs) and retain transcripts with orthologs in the sister taxon or in \Cgig (Thunberg, 1793).  We performed an all-by-all blastn \cite{BLAST} against \Espin {} or \Ecras {} at an E-value cutoff of $E<10^{-10}$.  We identified reciprocal best hits across the two species as orthologs. We identify 34,356 such direct 1:1 orthologs between \Espin {} and \Ecras. We estimate divergence across species using PAML \cite{PAML}. We pulled all fasta CDS sequences for direct orthologs, translated DNA to amino acids, aligned protein sequences using clustalw \cite{clustalw} then back translated to generate in-frame DNA alignments.  These in-frame alignments were analyzed in PAML codeml module \cite{PAML} to estimate dN and dS in pairwise mode.  \CgigFull, the marine Pacific oyster, is one of the best annotated outgroup genomes for freshwater mussels \cite{penaloza2021}.  Because \Cgig {} is at least 200 million years divergent from Unionids \cite{TimeTree,Bolotov2017}, we used a blastp \cite{BLAST} at an E-value cutoff of $E<10^{-20}$ to identify sequences with homology that likely indicate long-conserved functions. To identify putative TEs, we performed a tblastx \cite{BLAST} against the RepBase database \cite{RepBase} $E<=10^{-20}$. We also identified sequences with TE associated conserved domains such as reverse transcriptase, RNAse H, Integrase, gag, and any known retrotransposon or transposase related domains. After filtering we retain 32,679 putative genes (and pseudogenes) in \Espin {} and 34,763 genes (and pseudogenes) in \Ecras {} (Table \ref{Annot}).  

\subsubsection*{Gene Family Evolution} 
We estimated duplicate gene birth-death parameters \cite{Lynch2000} similarly to our prior work on mussels \cite{Rogers2021}.  To  identify gene families among annotations, we used a Fuzzy Reciprocal Best Hit Algorithm \cite{12Genomes} on an all-by-all blastn search of CDS sequences at $E<10^{-10}$ (self hits removed).  First order orthologs were retained and grouped into associated gene families.  For each gene family, we pulled all fasta CDS sequences, translated DNA to amino acids, aligned protein sequences using clustalw \cite{clustalw} then back translated to generate in-frame DNA alignments as previously described in prior work for \emph{M. nervosa} \cite{Rogers2021}.  These in-frame alignments were analyzed in PAML codeml module \cite{PAML} to estimate dN and dS across branches across the alignment guide tree.  Genes with $dN/dS>1.0$ on a branch after duplication are considered targets of positive selection. While reduced constraint can cause increased $dN/dS$ above the genomewide average, it is not expected to increase amino acid substitution SNPs ($dN$) above expectations based on neutral mutation rates ($dS$). These methods offer an advantage for non-model systems as they can be implemented using single genomes rather than requiring multi-genome phylogenomic panels.  

We observe 21 very large gene family expansions with $n >>100$ members per orthogroup  that contain protein domains that are sometimes but not always carried in TE or virus sequences (e.g. CHROMO domain, polymerases of putative viral origin, GIY-YIG domain).  We exclude these very large gene families with more than 100 members as putative viral or TE sequences except for 5 orthogroups where interpro domains clearly indicate cellular functions not related to selfish elements (e.g. lipid metabolism, drug and toxin resistance, immunity).  We used maximum likelihood to fit paramaters for birth $\lambda$, and death $\mu$ to describe rates of formation and loss of duplicate genes in natural populations. Using 1000 bootstrap replicates we established the 95\% CI for each parameter in whole genome data that could be used to establish significant departures from genomewide background dynamics. Whole genome mutation rates can be estimated from the birth parameter $\lambda$ per gene, scaled to the number of generations required to accumulate 1.0 $dS$ based on the mutation rate per nucleotide of $5\times10^{-9}$ \cite{Pogson1994}.  

\subsubsection*{Transposable elements}
We used RepDeNovo \cite{RepDeNovo} to identify recently proliferated transposable elements in the reference sample for \Espin {} and \Ecras. This program identifies repetitive reads from fastq files and de novo assembles repeat contigs independently from genome assembly or annotation so long as repeats are at copy number 10X or greater.  Repeat copies with less than 1.5-2\% sequence divergence typically cluster into a single consensus sequence, roughly equivalent to 2 million generations in mussels. Fastq sequences from independent sequencing runs were concatenated and used as input for RepDeNovo, run with default parameters.  Parameters for stimated coverage and genome size were taken from genome assembly metrics from Masurca \cite{masurca}.  Contigs assembled were classified into TE types using a tblastx against the RepBase database.    We mapped Illumina sequencing reads onto assembled repeat contigs to estimate copy number 

\subsubsection*{Heterozygosity and Population variation}
We mapped paired end Illumina data for the reference sample onto the respective reference using bwa aln and  sampe \cite{bwa}.  We identified SNPs in each sample using bcftools v1.1 mpileup and bcftools call -c.  SNPs were merged into a single vcf using bcftools merge.  We calculated sequencing depth using samtools depth.  We sequenced hDNA from 3 ethanol preserved samples graciously provided by Arthur Bogan and Jamie Smith at the North Carolina Museum of Natural Science: specimen numbers NC113372, NC27509, and NC30078.  We obtained performed phenol chloroform extractions to obtain hDNA, repaired ends with FFPE end repair, and then proceeded with a standard Illumina prep on unsheared DNA.  Samples were sequenced in multiplex on a NovaSeq lane. We obtained 6.0X coverage for NC113372 and 7.4X coverage for NC30078.  NC27509 had scarce DNA yield from tissues, and only produced 0.2X coverage and 90\% of sites had no coverage. It was excluded from downstream analyses.  Samples NC30078 and NC113372 were mapped to the reference genome using bwa mem \cite{BWAmem}  which performs better for partial alignments that are common in hDNA extracts.  

Small populations are prone to inbreeding as short coalescent times result in higher probability of relatedness across individuals \cite{Wright1922,Wright1949}, a key concern for conservation in this species with previously unknown genetic impacts \cite{us2011endangered,FiveYr}.  To identify putative regions of inbreeding, we counted the number of heterozygous sites per 10kb window across the genome.  We searched for genomic regions with full coverage but little to no heterozygosity. We also identified heterozygous sites genomewide at 4-fold synonymous and 4-fold nonsynonymous sites using CDS annotations, similarly to past work on \emph{M. nervosa} \cite{Rogers2023}.  Heterozygosity for \emph{M. nervosa}, \emph{E. hopetonensis}, and \emph{V. ellipsiformis} were generated using identical methods in previously published work \cite{Rogers2021}.  We do not offer similar estimates for \emph{Potamilus streckersonii} Smith, Johnson, Inoue, Doyle, and Randclev, 2019 that was also sequenced and assembled \cite{Smith2021} as the genome release includes long reads but not high quality short read Illumina data. As such these methods would not be appropriate for direct comparison. Similarly, we do not offer annotations for \emph{E. hopetonensis} as the Illumina only genome assembly is not sufficiently contiguous for such analysis at this time. 


\subsubsection*{Acknowledgements}
Illumina sequencing was performed at the Duke University Genome Sequencing Core and the Mt Sinai Sequencing core.  We thank Scott Small for specimen collection.  Art Bogan and Jamie Smith at the North Carolina Museum of Natural Science graciously provided \Espin {} historic specimens. The \Espin {} modern specimen for the reference genome was sampled under supervision of US Fish and Wildlife.  The UNC Charlotte High Performance Computing cluster supported computational analysis.  This work was funded on startup funds, a Faculty Research Grant, and other internal funding from UNCC. RLR is funded in part by  NIH NIGMS MIRA R35 GM133376.

\subsubsection*{Data availability}
Data will be released from the SRA and Dryad on acceptance.  

\subsubsection*{Author contributions}
RLR, JTG, and JPW designed experiments and analyses \\
JTG collected specimens from wild populations \\
RLR, JTG, and JPW wrote and edited the manuscript \\
\clearpage

\bibliography{Espinosa.bib}
\bibliographystyle{pnas}

\clearpage{}

\clearpage
%
%
%

\clearpage{}

\clearpage
\begin{figure}

A) \includegraphics[scale=0.4]{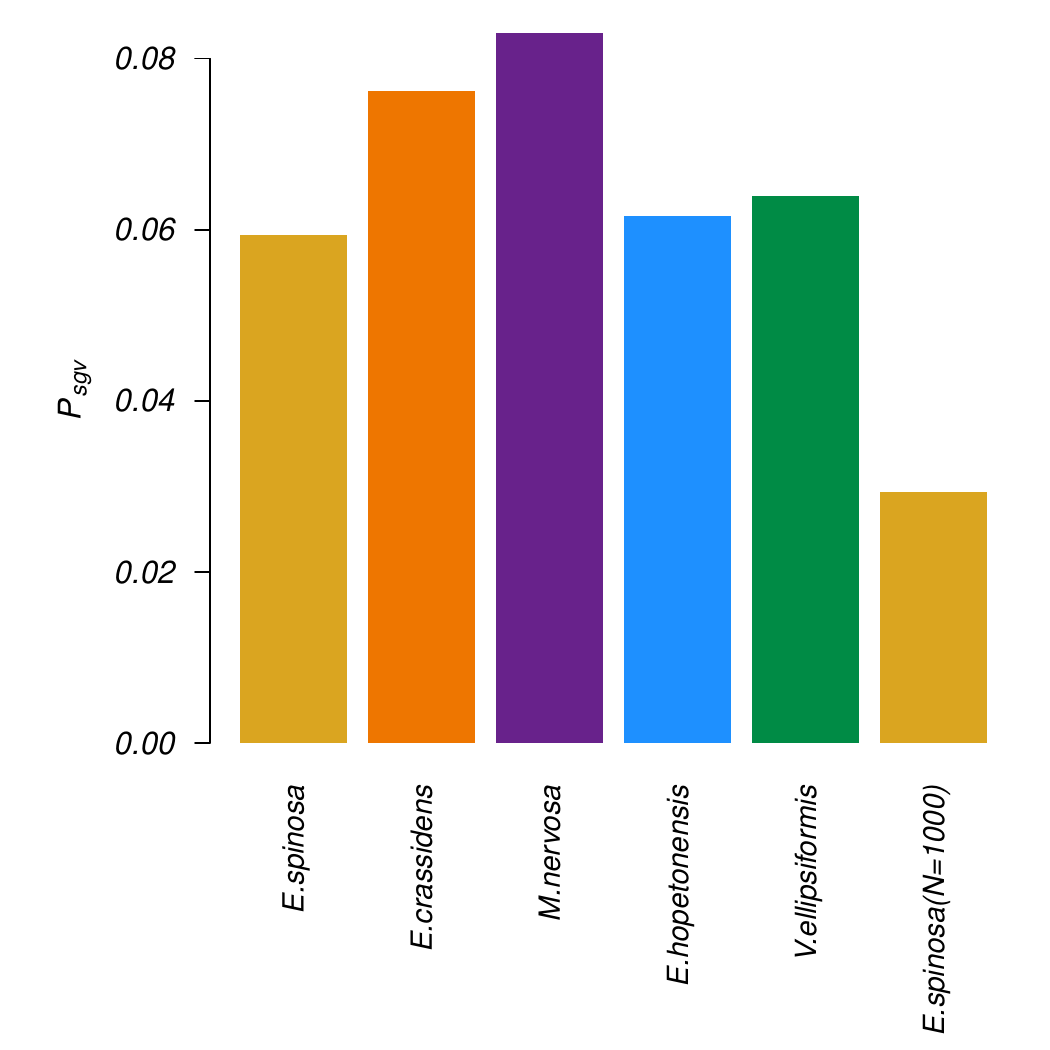}
B) \includegraphics[scale=0.4]{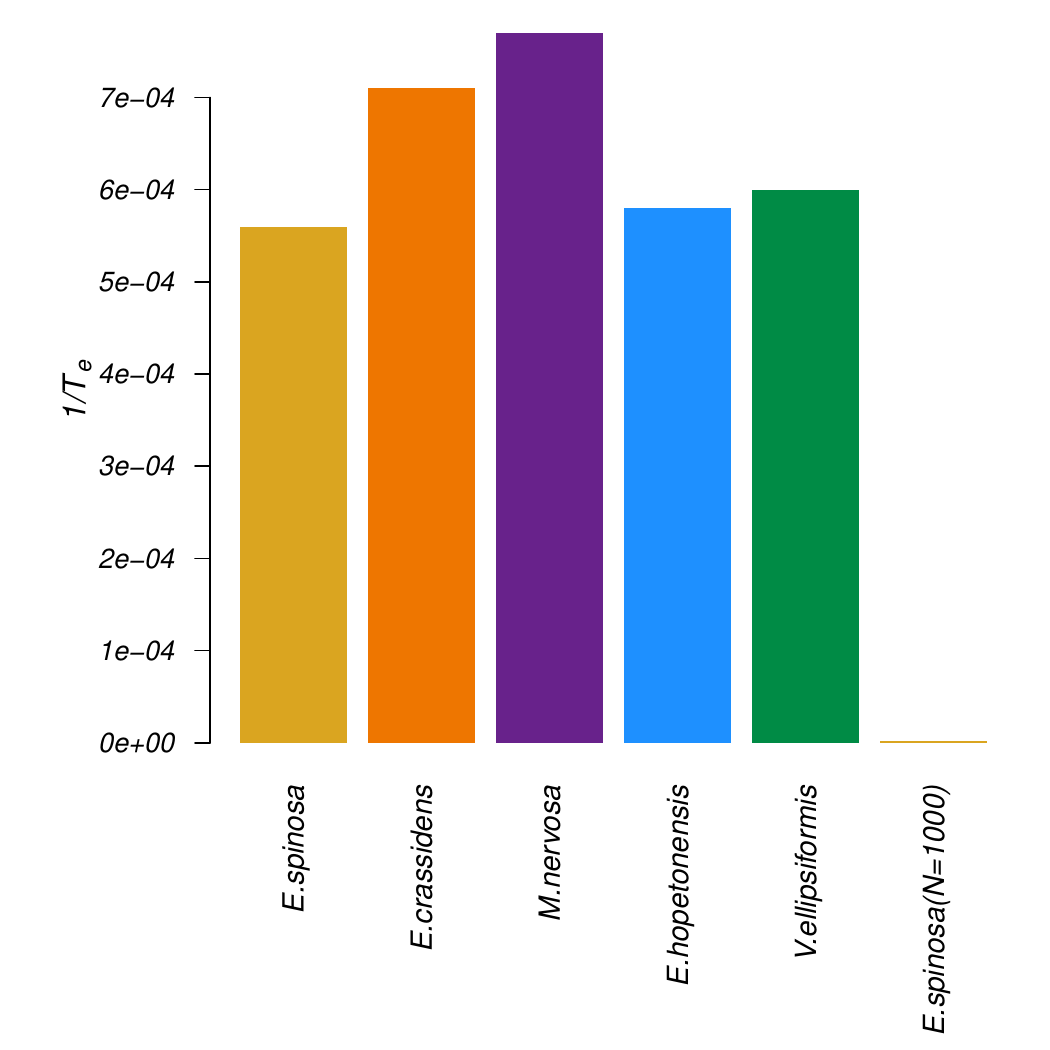}
\caption{Adaptive potential in freshwater mussels. Estimates of A) $P_{sgv}$ and B) $1/T_e$ in species of freshwater mussels, including theoretical estimates of $N=1000$ for endangered \emph{Elliptio spinosa}.  Under reduced population sizes the chances of adaptation via standing variation decline linearly. In contrast, the expected wait time for new mutations to appear and establish selective sweeps become prohibitively long and chances of new mutation per generation become low.   \label{Potential}}

\end{figure}

\clearpage{}

\clearpage{}

\begin{figure}
\includegraphics{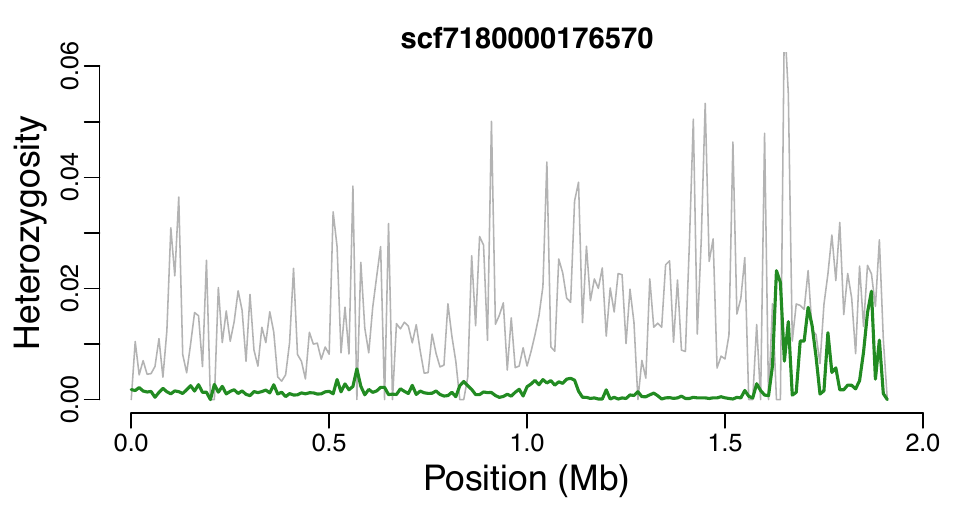}
\caption{Runs of homozygosity in \emph{E. spinosa} Reference sample on scf7180000176570 indicate inbreeding.  Near zero heterozygosity in the reference specimen (green) compared to hDNA for NC113372 (grey) is indicative of consanguineous mating.  Inbreeding tracts encompass 26.7\% of the genome, consistent with half-sibling matings. }
\end{figure}

\clearpage{}

\renewcommand{\thefigure}{S\arabic{figure}}
\renewcommand{\thetable}{S\arabic{table}}
\setcounter{figure}{0}
\setcounter{table}{0}
\renewcommand{\thetable}{S\arabic{table}}
\renewcommand{\thepage}{S\arabic{page}}
\setcounter{page}{1}
\section*{Supplementary Information}

\newpage

\newpage
\begin{table}
\caption{Genome Quality}
\begin{center}
\begin{tabular}{lrrr}
&  \emph{E. spinosa}  & \emph{Ecrassidens} \\
\hline
Genome Size (bp) &  2,857,182,358 & 3,485,727,846 \\
N50 (bp) &  285,217  & 232,187 \\ 
Coding sequence (bp) & 43,304,833 & 38,716,567 \\
\\
BUSCOs & \\
Single Copy & 860 (87.9\%)& 841 (86.0\%) \\
	 Duplicated & 43 (4.4\%) &  65 (6.6\%) \\
	 Fragmented & 32 (3.3\%) & 33 (3.4\%) \\
	 Missing & 43 (4.4\%) &  39 (4.0\%)\\
	 Total Searched & 978 & 978 \\
	 \hline
\end{tabular}
\end{center}
\end{table}

\clearpage

\begin{table}
\caption{Genome Annotations \label{Annot}}
\begin{center}
\begin{tabular}{lrr}
& \emph{E. spinosa} & \emph{E. crassidens} \\
\hline
Gene Sequences & 32,679  & 34,673 \\
CDS Sequences & 33,630  & 39,964 \\
Mean Gene Length & 17,070 & 18,648   \\
Mean CDS Length & 1347 & 1194 \\
Mean Intron Length & 4092 & 4324 \\
Mean Exon Length & 269 & 212 \\
\hline
\end{tabular}
\end{center}
\end{table}

\clearpage
\begin{table}
\caption{Heterozygosity and Constraint \label{HetTable}}
\begin{center}
\begin{tabular}{llrrrr}
Species & Sample & $H_0$ & $H_C$ $^1$ & $H_N/H_S$   \\
\hline
 \emph{E. spinosa} & Reference & 0.0041 & 0.0056 & 0.40 \\
\emph{E. spinosa} & NC30078 & 0.0125  & 0.0132 & 0.44   \\
\emph{E. spinosa} & NC 113372 &0.0139  & 0.0148 & 0.53  \\
 \emph{E. crassidens} & Reference & 0.0071 & 0.0071 & 0.32 & \\
 \emph{E. hopetonensis}& Reference & 0.0058 & - & -  \\
 \hline
\emph{M. nervosa}& Reference & 0.0077 &  - & - \\
\emph{V. ellipsiformis}& Reference & 0.0060 & - & -  \\
\hline
\end{tabular}
\end{center}
$^1$ Corrected for inbreeding \\
\end{table}

\clearpage


\begin{table}
\caption{Gene Duplication Rates in \emph{E. spinosa} \label{DupRateEspin}}
\begin{center}
\begin{tabular}{lrrrrl}
Gene Family & \# genes & $\lambda_{O}$ & $\lambda_{E}$  & $\mu_{O}$  \\
\hline
Whole Genome & 32,679 & 119,416 & - & 11.1 $\pm$0.1 \\
Cytochromes & 183 & 1,927 & 669 & 11.4  \\
ABC Transporters & 124 & 1,987 & 453 & 11.9   \\
von Willebrand Factors & 189 & 1,372 & 691 & 10.8 \\
Fibrinogens & 220 & 2,835 & 804 & 13.5  \\
Opsins & 161 & 392 & 588 & 10.6  \\
Mitochondria Eating & 168 & 1,021 & 614 &7.5 \\
Hsp70  & 153 & 1,104 & 559 & 8.9  \\
\hline
\end{tabular}
\end{center}
$\lambda_O$ Observed Birth Rate per gene in 1.0 $dS$ \\
$\lambda_E$ Expected Birth Rate per gene in 1.0 $dS$ based on whole genome dynamics \\
$\mu_O$ Observed Death Rate per gene in 1.0 $dS$ \\
\end{table}

\clearpage

\begin{table}
\caption{Gene Duplication Rates in \emph{E. crassidens} \label{DupRateEcras}}
\begin{center}
\begin{tabular}{lrrrl}
Gene Family & \# genes & $\lambda_{O}$ & $\lambda_{E}$  & $\mu_{O}$ \\
\hline
Whole Genome & 34,673& 87,661 & - & 8.5 $\pm$0.1  \\
Cytochromes & 138 & 673 & 349 & 8.2  \\
ABC Transporters & 112 & 706 & 283 & 7.2  \\
von Willebrand Factors & 233 & 1675 & 589 & 9.2   \\
Fibrinogens & 250 & 1326 & 632 & 8.5  \\
Opsins & 199 & 605 & 503 & 8.9  \\
Mitochondria Eating & 185 & 642 & 468 & 6.1 \\
Hsp70  & 210 & 1601 & 531 & 8.7 \\
Chitin & 69 & 411 & 174 & 7.9  \\
\hline
\end{tabular}
\end{center}
$\lambda_O$ Observed Birth Rate per gene in 1.0 $dS$ \\
$\lambda_E$ Expected Birth Rate per gene in 1.0 $dS$ based on whole genome dynamics \\
$\mu_O$ Observed Death Rate per gene in 1.0 $dS$  \\

\end{table}

\clearpage

\begin{table}

\caption{Adaptive Gene Duplications ($dN/dS >>1.0$) \label{DupSelnTable}}
\begin{center}
\begin{tabular}{lrr}
Functional Class & \emph{E. spinosa} & \emph{E. crassidens} \\
\hline
ABC Transporters & 10 &	12 \\
Cytochromes & 18 & 17\\
Fibrinogens & 35	& 29\\
Hsp70 & 25 & 36 \\
Mitochondria Eating Proteins & 25 & 35 \\
Opsins & 11 & 11 \\
VonWillebrand Proteins & 14 & 40 \\
\hline
Total & 138 & 180 \\
Adjusted$^1$  & 103 & 180 \\
\hline
\end{tabular}
\end{center}
\footnotesize$^1$ Adjusted for genomewide background duplication rate

\end{table}

\clearpage

\begin{table}
\caption{Transposable Element Content (Mb) \label{TETable}}
\begin{center}
\begin{tabular}{lrrrrrr}
TE Type & \emph{E. spinosa} & \emph{E. crassidens} & \emph{E. hopetonensis} & \emph{M. nervosa} & \emph{V. ellipsiformis}\\
\hline
Polinton  & 11.69 & 68.80 & 25.00 & 49.17 & 99.60  \\ 
Gypsy & 5.07 & 49.34 & 16.50 & 40.50 & 16.20 \\ 
RTE & 2.79 & 20.82 & 5.85 & 11.70 & 4.64 \\ 
hAT & 2.73 & 8.18 & 1.69 & 2.09 & 0.08\\ 
Crack  & 1.16 & 19.12 & 4.09 & 5.26 & 4.15\\ 
LINE & 0.95  & 21.11 & 7.38 & 16.11 & 4.33\\ 
Penelope & 0.92 & 9.40 & 2.60 & 3.50 & 3.29\\  
Mariner  & 0.54 & 2.83 & 1.29 & 1.44 & 1.33 \\ 
Neptune & 0.14 & 0 & 0 & 6.21 & 0 \\ 
Unclassified & 75& 511 & 123 & 250 & 13\\
\hline
Total  & 101 & 711 & 188 & 386 & 148  \\
\end{tabular}
\end{center}
\end{table}

\clearpage

\begin{table}
\caption{Adaptive Potential \label{PotentialTable}}
\begin{center}
\footnotesize
\begin{tabular}{|lrr|cr|crr| }
\hline
 & \multicolumn{2}{c}{} &\multicolumn{2}{c}{SNPs } &  \multicolumn{2}{c}{ Duplicates} &  \\
Species  & $H_C$ & $N_e$ & $T_e$ (gens) &  $P_{sgv}$  & Dup Rate & $T_e$ (gens) & $P_{sgv}$    \\
\hline
 \emph{E. spinosa}  &  0.0056 &280,000 & 1786  & 0.0594 &  $2.53\times10^{-8}$ & 245 &0.3604 \\
 \emph{E. crassidens}  & 0.0071 & 355,000 & 1408  & 0.0762  &$3.65\times10^{-8}$ & 278 &  0.3306\\
\emph{M. nervosa}& 0.0077 &  385,000& 1299 & 0.0830  & $1.16\times10^{-8}$ & 560  & 0.1821 \\
  \emph{E. spinosa} (N=1000)  &  0.0056 & 1,000 & 500,000 &  0.0293 & $2.53\times10^{-8}$ & 68,493 & 0.1949\\
 \hline
 \emph{E. hopetonensis}&  0.0058 & 290,000 &  1724 & 0.0616 & - & - & - \\
\emph{V. ellipsiformis}&  0.0060 & 300,000 & 1667 & 0.0639& - & - & - \\
\hline
\end{tabular}
\end{center}
  Mutation rate $\mu=5\times10^{-9}$ for SNPs. \\
 Assume selection coefficient $s=0.1$. \\
\end{table}
\clearpage

\end{document}